\documentclass[twocolumn,showpacs,nofootinbib]{revtex4}
\usepackage{graphicx}
\usepackage{dcolumn}
\input epsf
\def\be{\begin{equation}}
\def\ee{\end{equation}}
\def\bea{\begin{eqnarray}}
\def\eea{\end{eqnarray}}

\begin{document}
 \tighten
\parskip 4pt
\renewcommand{\topfraction}{0.8}
\vskip 3cm

\

\preprint{SU-ITP-03/01,\,  SLAC-PUB-9630, \,TIFR/TH/03-03}
\title {\Large\bf Landscape, the Scale of SUSY Breaking, and Inflation}
 \author{\bf Renata Kallosh and Andrei Linde}
\address{ {Department
  of Physics, Stanford University, Stanford, CA 94305-4060,
USA}    } {\begin{abstract} We argue that in the simplest version of the KKLT
model, the maximal value of the Hubble  constant during inflation cannot exceed
the present value of the gravitino mass, $H\lesssim m_{3/2}$. This may have
important implications for string cosmology and for the scale of the SUSY
breaking in this model. If one wants to have inflation on high energy scale,
one must develop phenomenological models with an extremely large gravitino
mass. On the other hand, if one insists that the gravitino mass should be O(1
TeV), one will need to develop models with a very low scale of inflation. We
show, however, that one can avoid these restrictions in a more general class of
KKLT models based on the racetrack superpotential with more than one exponent.
In this case one can combine a small gravitino mass and low scale of SUSY
breaking with the high energy scale of inflation.
\end{abstract}}
 \maketitle
\section{Introduction}

Soon after the invention of inflationary cosmology, it became clear that our universe may consist of many exponentially large locally homogeneous regions corresponding to different stable or metastable vacuum states, which could be used for justification of anthropic principle \cite{Linde:1984je}. It was argued that the total number of such states obtained as a result of compactification of 10D or 11D universe can be exponentially large \cite{Sakharov:1984ir}. Early estimates of the total number of different vacua in heterotic string theory gave astonishingly large numbers such as $10^{1500}$ \cite{Lerche}. A more recent investigation, based on the idea of flux compactification, gave similarly large number of possible vacua \cite{Bousso}.

However, none of the stable or metastable 4D vacua  which were known in string theory at that time described dS space or accelerating/inflationary universe. Some progress in finding metastable dS vacua  with a positive cosmological constant was achieved only recently \cite{KKLT} (see also \cite{Silver}, where this problem was addressed for noncritical string theory). The main idea of KKLT was to find a supersymmetric AdS minimum taking into account nonperturbative effects, and then uplift this minimum to dS state by adding the positive energy density contribution of $\overline{\rm D3}$ branes or D7 branes. The position of the dS minimum and the value of the cosmological constant there depend on the quantized values of fluxes  in the bulk and  on the branes. This provided a starting point for a systematic investigation of the landscape of all possible metastable dS string theory vacua
\cite{Susskind:2003kw, Douglas:2003um}. 

The idea of string theory landscape may be useful, in particular, for understanding the scale of supersymmetry breaking. Different points of view on this issue have been expressed in the literature, with the emphasis on  statistics of string theory flux vacua \cite{Susskind:2004uv,Douglas:2004qg,Dine:2004is,Conlon:2004ds,Kachru}. Depending on various assumptions, one can either conclude that most of the vacua in the string theory landscape correspond to strongly broken supersymmetry, or find a large set of vacua with a small-scale SUSY breaking.  The methods used in \cite{Susskind:2004uv,Douglas:2004qg,Dine:2004is,Conlon:2004ds,Kachru} study the distribution of  flux vacua and do not involve the study of the shape of the effective potential after the uplifting to dS state. The relation between the parameters of SUSY breaking, the height of the barrier stabilizing dS vacua and the Hubble constant during inflation was not studied so far. This will be the main subject of our investigation.

First of all, we will show that the gravitino mass in the KKLT scenario  with
the superpotential used in \cite{KKLT} is extremely large, $m_{3/2} \sim
6\times 10^{10}$ GeV. This means that in the context of this model one should
either consider particle phenomenology with superheavy gravitino, see e.g.
\cite{DeWolfe:2002nn, Arkani-Hamed:2004fb}, or modify the scenario in such a
way as to make it possible to reduce the gravitino mass by many orders of
magnitude.

We will show also that the Hubble parameter during inflation in the simplest
models based on the KKLT scenario cannot exceed the present value of the
gravitino mass, $H \lesssim m_{3/2}$. This means that even if we succeed to
find the models  with small gravitino mass, following
\cite{Dine:2004is,Conlon:2004ds,Kachru}, we may face an additional problem of finding
successful inflationary models with extremely small H.

We will then suggest a possible resolution of these problems in the context of
a volume modulus stabilization model where the gravitino mass is not related to
the scale of inflation and can be made arbitrarily small.

\section{Gravitino-Hubble relation in the simplest KKLT model}

Recent ideas on string cosmology rely on a possibility to stabilize string
theory  moduli, in particular the dilaton and the total volume modulus. The
simplest KKLT models \cite{KKLT} with the superpotential of the form $W=W_0+
Ae^{-a\rho}$ and with the K\"ahler potential $K = - 3 \ln[\rho +
\overline{\rho}]$  provide the AdS minima for the volume modulus $\rho =
\sigma +i\alpha$ at finite, moderately large values of volume.  When this
potential is supplemented by a D-type  contribution $C\over \sigma^2$ from
$\overline{\rm D3}$ brane \cite{KKLT} or D7 branes \cite{BKQ}, one finds a de Sitter
minimum. This simplest KKLT model has a minimum at some real value of the field
$\rho$: $\rho = \sigma$, $\alpha = 0$.  This minimum is separated from the
Minkowski vacuum of Dine-Seiberg type at infinite volume of the internal space
by a barrier, which makes the de Sitter minimum metastable with the lifetime $t
\sim 10^{10^{120}}$ years.
\begin{figure}[h!]
\centering\leavevmode\epsfysize=5.5cm \epsfbox{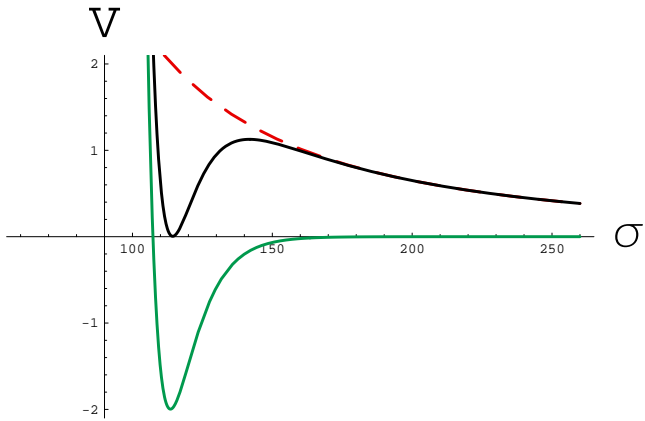} \caption[fig1]
{Thin green line corresponds to AdS stabilized potential for $W_0 =- 10^{{-4}}$,
$A=1$, $a =0.1$. Dashed line shows the additional term $C\over \sigma^2$,
which appears either due to the contribution of a $\overline{D3}$ brane or of a
D7 brane. Thick black line shows the resulting potential including the $C\over
\sigma^2$ correction with $C = 2.6\times 10^{{-11}}$, which uplifts the AdS minimum to a
dS minimum. All potentials are shown multiplied by $10^{15}$.} \label{1}
\end{figure}

Since $D_i W
 =0$ in the AdS minimum, its  depth is given by
 \be\label{ads}
V_{\rm AdS} ~=~ -3 e^{K}|W|^2 \ . \label{fluxsup} \ee Here all functions are
calculated  at $\sigma =\sigma_{\rm cr}$,  where $\sigma_{\rm cr}$ is the
position of the minimum of the potential prior to the uplifting. We use the
units where $M_P^2=(8\pi G_N)^{-1}=1$.

Before the uplifting, the potential has only one extremum, at $\sigma=
\sigma_{\rm cr}$, and its absolute value exponentially decreases at $\sigma \gg
\sigma_{\rm cr}$. When we add the term ${C\over \sigma^{2}}$, the minimum
shifts upward in such a way that the new dS minimum is positioned at
$\sigma_{0} \approx \sigma _{\rm cr}$. This means that the values of the
function $e^{K(\sigma)} |W(\sigma)|^2$  in the minimum of the effective
potential remain almost unchanged during the uplifting. Meanwhile, the value of
$D_i W(\sigma)$ in the minimum after the uplifting is no longer equal to zero,
but it  still remains relatively small, $D_i W(\sigma_{0}) \ll W(\sigma_{0})$.
At the dS minimum,  the total effective potential must vanish, with the
accuracy of $10^{-120}$. Therefore one has  ${C\over \sigma_{0}^{2}} \approx
-V_{\rm AdS} ~=~ 3 e^{K}|W|^2 $.

The gravitino mass in the uplifted dS minimum is given by
  \be\label{gravitino1}
m^{2}_{3/2}(\sigma_{0}) ~=~ e^{K(\sigma_{0})}|W(\sigma_{0})|^2 \approx
e^{K(\sigma_{\rm cr})}|W(\sigma_{\rm cr})|^2=  {V_{\rm AdS}\over 3}\ .
\label{fluxsup} \ee

The gravitino mass can be associated with the strength of supersymmetry
breaking at the minimum where the total potential is approximately vanishing.
Indeed, \be V_{\rm KKLT}(\sigma_{0})= V_F+V_D= |F|^2- 3 m_{3/2}^2 + {1\over
2}D^2 \approx 0 \ . \ee
 This yields
\be
 3 m_{3/2}^2 \approx  {1\over 2}D^2+|F|^2 \ .
\ee

Now let us discuss the height of the barrier $V_{\rm B}$ which stabilizes  dS
state after the uplifting. Since the uplifting is achieved by adding a slowly
decreasing function $C/\sigma^{2}$ to a potential which rapidly approaches zero
at large $\sigma$,  the height of the barrier $V_{\rm B}$ is approximately
equal (up to a factor $O(1)$)  to the depth of the AdS minimum $V_{\rm AdS}$,
see Fig. \ref{1}:
 \be\label{gravitino2}
V_{\rm B}\sim  |V_{\rm AdS}| \sim  m^{2}_{3/2} \ .
\label{fluxsup}
\ee

To complete the list of important features of this model, let us remember what should be done to use it for the description of inflation.

The simplest possibility would be to use the extremum of the potential of the height $V_{\rm B}$ as an initial point for inflation. A particular realization of this scenario  was proposed in \cite{Blanco-Pillado:2004ns}. (In order to do it, it was necessary to consider a racetrack superpotential with two exponents). In this case one has an interesting relation between various parameters of our model and the Hubble constant during inflation:
 \be\label{gravitino3}
H^{2}\approx V_{\rm B}/3 \sim  |V_{\rm AdS}|/3 \sim  m^{2}_{3/2} \ .
\label{fluxsup}
\ee

\begin{figure}[h!]
\centering\leavevmode\epsfysize=5.5cm \epsfbox{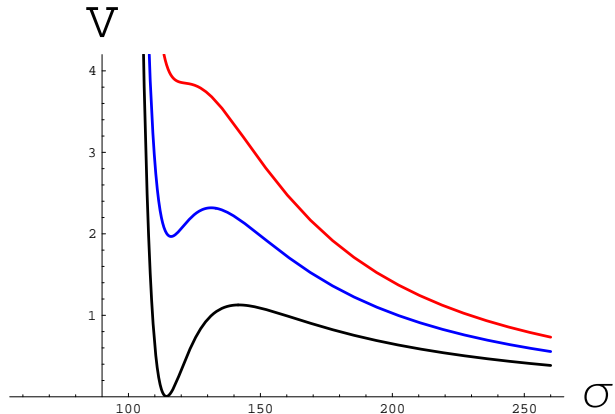} \caption[fig2]
{The lowest curve with dS minimum is the one from the KKLT model. The second one describes, e.g., the D3/D7 inflationary potential with the term $V_{\rm infl}={V(\phi)\over \sigma^3}$ added to the KKLT potential; it originates from fluxes on D7 brane. The top curve shows that when the  inflationary potential becomes too large, the barrier disappears, and the internal space decompactifies. This explains the constraint $H\lesssim   m_{3/2}$.  } \label{2}
\end{figure}

One may also achieve inflation by considering dynamics of branes in the compactified space. This involves a second uplifting, which corresponds to a nearly dS (inflationary) potential added to the KKLT potential $V_{KKLT}$, for example in D3/D7 case \cite{KKL}. The added potential should be flat in the inflaton direction, and, according to  \cite{KKL}, it has a $\sigma^{-3}$ dependence   on the volume modulus:
\be
V_{\rm tot}^{\rm infl} \approx V_{KKLT} (\sigma) + {V(\phi)\over \sigma^3} \ .
\ee
Here $\phi$ is an inflaton field. The resulting potential as a function of $\sigma$ is schematically shown in Fig. \ref{2} for different values of the function $V(\phi)$. It is apparent from this figure that the vacuum stabilization is possible in this model only for sufficiently small values of the inflaton potential,
\be \label{gravitino4}
V_{\rm tot}^{\rm infl} \lesssim c\ V_{\rm B} \sim c\ |V_{\rm AdS}| \sim  c\  m^{2}_{3/2} \ ,
\ee
where $c \approx 3$ for the original version of the KKLT model.\footnote{This effect is similar to decompactification of space at large $H$ studied in \cite{Linde:1988yp}, and to the dilaton destabilization at high temperature discussed in \cite{Buchmuller:2004xr} in a different context. A related effect was also found in \cite{Gen:2000nu,Frolov:2003yi} and  \cite{Navarro:2004mm} in models of radion stabilization. }

The key reason for the vacuum destabilization is the  $\sigma^{-n} $  dependence of the inflaton potential, with $n>0$ .  As explained in  \cite{Giddings,KKLMMT}, the   runaway $\sigma^{-n}$ dependence of the energy density in string theory is quite generic. The $\sigma^{-3}$ dependence  appears explicitly in the D-term contribution to the vacuum energy, which is the source of the inflationary potential $V(\phi)/\sigma^{3}$ in D3/D7 inflation \cite{KKL}.   In principle, it might be possible to design inflationary models where the inflaton potential depends on $\sigma$ and $\phi$ in a more complicated way due to some nonperturbative effects involving both fields. This could prevent vacuum destabilization at large energy density. However, no examples of such models  are known.

Equations (\ref{gravitino3}), (\ref{gravitino4}) provide a strong constraint on the Hubble constant during inflation in a broad class of KKLT-based inflationary models:
\be\label{gravitino5}
H\lesssim   m_{3/2} \ .
\label{fluxsup}
\ee

One should note that there could be many stages of inflation in the early universe, some of which could happen in a vicinity of a different minimum of the effective potential in stringy landscape, with much higher  barriers surrounding it. Thus it is quite possible that at some stage of the evolution of the universe the Hubble constant was much greater than $m_{3/2}$.  However, this could not be the last stage of inflation. We cannot simply jump to the KKLT minimum after the tunneling with bubble formation following some previous stage of inflation, because such tunneling would create an open universe. After such tunneling, we will still need to have a long stage of inflation, which should make the universe flat, form the large scale structure of the observable part of the universe, and end by a slow roll to the KKLT minimum. Our results imply that the Hubble constant $H$ at this last and most important stage of inflation  should be smaller than the present value of the gravitino mass.

\section{Problems with SUSY breaking and inflation in the simplest KKLT model}

Now we are ready to formulate a list of unusual features of this scenario.

1) If one takes the simplest superpotential of the KKLT model according to \cite{KKLT}, one finds, following (\ref{ads}), (\ref{gravitino1}), that  the gravitino mass in this scenario is extremely large, $m_{3/2} \sim  \sqrt{V_{\rm AdS}} \sim 2.5 \times 10^{-8} M_{p} \sim 6\times 10^{{10}}$ GeV. Other parameters characterizing the strength of supersymmetry breaking have similar magnitude. These numbers are many orders of magnitude higher than the gravitino mass O(1 TeV) often discussed in the literature.

In this situation there are two basic choices. The first idea that comes to mind is to change the parameters of the KKLT model in such a way as to reduce the scale of SUSY breaking and the gravitino mass down to the TeV scale. This is a rather nontrivial task, which is the subject of many recent investigations \cite{Susskind:2004uv,Douglas:2004qg,Dine:2004is,Kachru}. Our results add two new problems to the list of the problems studied in these papers.

First of all, if our observations based on the simplest KKLT model are generic (this question is not addressed by the methods of Refs. \cite{Susskind:2004uv,Douglas:2004qg,Dine:2004is,Conlon:2004ds,Kachru}), then the minimum of the KKLT potential is extremely shallow, with the low barrier height $V_{\rm B} \sim    m^{2}_{3/2} \lesssim 10^{-30}$ in Planck density units. This implies that one should be especially careful when analyzing the possibility that the field $\sigma$ during its cosmological evolution may overshoot the KKLT minimum and roll over the barrier, which will lead to decompactification of the 4D space \cite{Brustein:1992nk}; see \cite{kaloper,Blanco-Pillado:2004ns} for a list of proposed solutions of this problem.

Another problem is that we will need to find inflationary models with $H \lesssim 1$ TeV, i.e. with $H \lesssim 10^{{-15}}$ in Planck units. Whereas such models can be quite satisfactory from the cosmological point of view, no explicit examples of models of such type have been constructed so far in string theory with stable internal dimensions.

Another option  is to develop particle phenomenology based on the models with extremely large scale of SUSY breaking. This is a very interesting possibility, which was recently discussed, e.g.,  in \cite{DeWolfe:2002nn, Arkani-Hamed:2004fb}.

In this paper we are going to suggest a different route which may help us to solve the problems discussed above.

\section{New Features in the Landscape:  Supersymmetric Minkowski Vacua and Light Gravitino}

The problems discussed above are related to the fact that the simplest  KKLT potential has only one minimum, and this minimum occurs at large negative values of the effective potential.  Therefore we will look  for a possibility to stabilize the volume modulus in a supersymmetric Minkowski minimum.
We perform an analysis of the vacuum structure\footnote{We performed the calculations and we plot the corresponding potentials using the ``SuperCosmology'' code \cite{S}.} keeping the tree-level K\"ahler
potential
$
K = - 3 \ln[(\rho + \overline{\rho})]
$
and a  racetrack superpotential similar to the one recently used in the  racetrack
inflation scenario \cite{Blanco-Pillado:2004ns}
\be
W = W_0 + Ae^{-a\rho}+ Be^{-b\rho} \ .
\label{adssup}
\ee
Here $W_0$ is a tree level contribution which arises from the fluxes. The exponential terms
arise  either from Euclidean D3 branes of from gaugino condensation  on D7 branes, as explained in
\cite{KKLT,Blanco-Pillado:2004ns}.

\begin{figure}[h!]
\centering\leavevmode\epsfysize=7cm \epsfbox{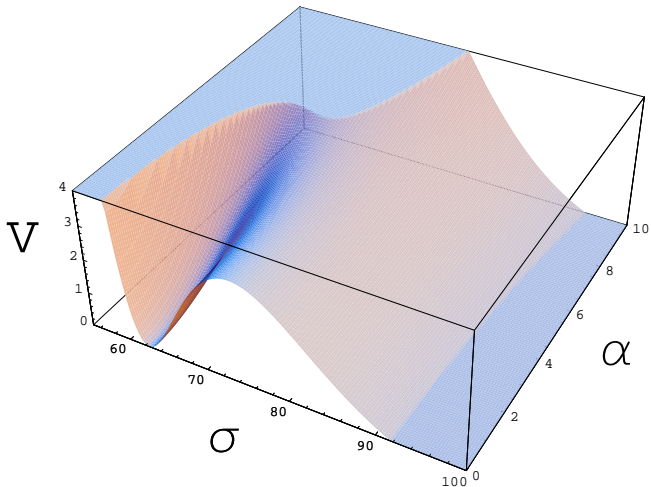} \caption[fig2]
{The potential as a function of the complex field $\rho$. The Minkowski minimum
occurs at $\alpha = {\rm Im}~\rho =0$, as we have assumed in the analytic
investigation.} \label{4}
\end{figure}

At a  supersymmetric vacuum  $D_\rho W=0$.
The supersymmetric Minkowski  minimum then lies at
\be
W(\sigma_{cr})=0 \ , \qquad DW(\sigma_{cr})=0 \ .
\label{susy} \ee
As in KKLT, we simplify things by setting the imaginary part of
 the $\rho$ modulus (the axion field $\alpha$) to zero, and
letting $\rho =\overline{\rho}= \sigma$. (Even though in some models the condition $\alpha = 0$  is not satisfied at the minimum of $V(\rho)$ \cite{Blanco-Pillado:2004ns}, we have verified that it is satisfied in the model which we are going to propose; see Fig. \ref{4}.) In addition we take  $A,a$, $B, b$ and $W_0$ to be  all real
and the sign of $A$ and $B$ opposite.

We find a simple relation between the critical value of the volume modulus  and
parameters of the superpotential \be
 \sigma_{cr}= {1\over a-b}\ln \left |{a\,A\over b\,B}\right|\, .
\label{sigmacr} \ee Equations (\ref{susy}) require also a particular relation
between the  parameters of the superpotential: \be -W_0= A \left |{a\,A\over
b\,B}\right|^{a\over b-a} +B \left |{a\,A\over b\,B}\right| ^{b\over b-a} \ee
Note that only solutions with non-vanishing value of $W_0$ are possible in this
model; these solutions disappear if we put $A$ or $B$ equal to zero, as in the
original version of the KKLT model.

The potential, $V= e^{K}\left(  G^{\rho\overline\rho} D_{\rho}W \overline {D_{\rho} W}
- 3|W|^2 \right)$, as the function of the real field $\rho =\overline{\rho}= \sigma$ is given by
\bea
&&V= {e^{-2(a+b)\sigma}\over 6\sigma^2}(b B e^{a\sigma}+ a A e^{b\sigma})\nonumber
\\ &\times& \left [Be^{a\sigma}(3+b\sigma)+e^{b\sigma}(A(3+a\sigma)+3e^{a\sigma}W_0)\right]
\label{pot}\eea It vanishes at the minimum which corresponds to Minkowski
space: \be V_{\rm Mink}(\sigma_{cr})=0 \ ,  \qquad {\partial V\over \partial
\sigma}(\sigma_{cr})=0 \ . \ee Thus it is possible to stabilize the volume
modulus while preserving  Minkowski supersymmetry. The gravitino mass in this
minimum vanishes.

\begin{figure}[h!]
\centering\leavevmode\epsfysize=5.3cm \epsfbox{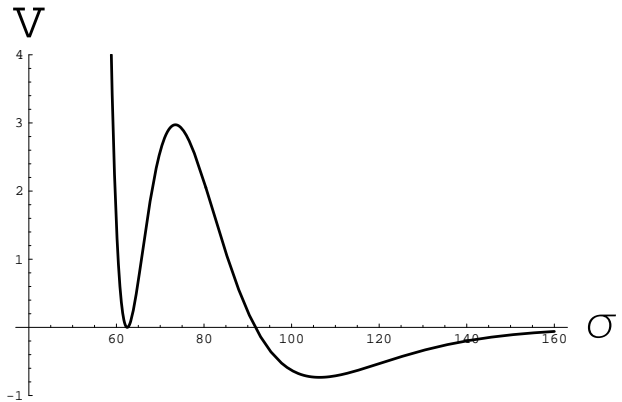} \caption[fig2]
{The F-term potential (\ref{pot}), multiplied by $10^{14}$, for the values of the parameters $A=1,\ B=-1.03,\ a=2\pi/100,\ b=2\pi/99,\ W_0= - 2\times 10^{-4}$. A Minkowski minimum at $V=0$ stabilizes the volume at $\sigma_{cr}\approx 62$. AdS vacuum at $V<0$ stabilizes the volume at $\sigma_{cr}\approx 106$.  There is a barrier protecting the Minkowski minimum. The height of the barrier is not correlated with the gravitino mass, which   vanishes  if the system is trapped in Minkowski vacuum. } \label{3}
\end{figure}

An example of the model where the vacuum stabilization occurs in the
supersymmetric Minkowski vacuum is given by the theory with the superpotential
(\ref{adssup}) with $A=1,\ B=-1.03,\ a=2\pi/100,\ b=2\pi/99,\ W_0= -2\times
10^{-4}$. None of these parameters is anomalously large or small; they are of
the same order as  the parameters used in \cite{KKLT}. The resulting potential
is shown in Figs. \ref{3} and \ref{4}. The vacuum stabilization occurs at
$\sigma \approx 62 \gg 1$, which suggests that the effective 4D supergravity
approach used in our calculations should be valid.

We have found the supersymmetric Minkowski vacuum prior to adding any
nonperturbative terms  $\sim C/\sigma^{2}$ related to $\overline{\rm D3}$ brane or D7
branes.   We assume, as usual, that by changing the parameters and by adding
the term $C/\sigma^{2}$ one can fine-tune the value of the potential in its
minimum to be equal to the observed small constant $\Lambda \sim 10^{{-120}}$.
What is important for us is that in the first approximation one can make the
gravitino mass vanish as compared  to all other parameters of the
superpotential. As a result, the value of $m_{3/2}$ in our model does not have
any relation to the height of the potential, and, correspondingly, to the
Hubble constant during inflation.

An important property of our Minkowski (or near-Minkowski) vacuum, as well as
the dS vacuum obtained by its uplifting, is that the gravitino mass vanishes
(or nearly vanishes) only in its vicinity. Similarly, restoration of
supersymmetry in this minimum implies that all particles whose mass is
protected by supersymmetry are expected to be light in the vicinity of the
minimum. However, supersymmetry breaks down and all of these particles become
heavy once one moves away from the minimum of the effective potential. This is
exactly the situation required for the moduli trapping near the enhanced
symmetry points according to \cite{Kofman:2004yc} (see also  \cite{McAllister:2004gd,W}). This
suggests that the moduli trapping may help us to solve the overshooting problem
in our scenario. The fact that the minimum of the effective potential is
simultaneously a trapping point is a distinguishing feature of our model.

The same model may also have AdS vacua defined by
\be
W(\sigma)\neq 0\ , \qquad DW(\sigma)= 0 \ .
\label{AdS} \ee

At the AdS minimum  one has \be -W_0= Ae^{-a\sigma}(1+ {2\over 3}a\sigma) +
Be^{-b\sigma}(1+ {2\over 3}b\sigma) \ . \ee The vacuum energy in this minimum is
negative, 
\be V(\sigma)=-3e^K|W|^2 =- {(a A e^{-a\sigma} +b B
e^{-b\sigma})^2\over 6\sigma} \ . 
\ee 
In our particular example shown in Figs.
\ref{3} and \ref{4} the AdS  minimum occurs at $\sigma \sim 106$.

The supersymmetric Minkowski vacuum is absolutely stable with respect  to the
tunneling to the vacuum with a negative cosmological constant. Indeed,
tunneling from a supersymmetric Minkowski vacuum would require creation of
bubbles of a new phase with vanishing total energy, which is impossible because
of the positive energy theorems \cite{Weinberg:1982id}.

This state may become metastable after the uplifting of the Minkowski minimum
(or of a shallow AdS minimum) to the dS minimum with $\Lambda \sim
10^{{-120}}$. Since the tunneling  will occur through the barrier with mostly
positive $V(\sigma)$, one would expect that the lifetime of the dS space will
be about $10^{10^{{120}}}$ years, as in the simplest KKLT model \cite{KKLT}.
However, this question requires a more detailed investigation. An important
distinction of the tunneling in the simplest  model of Ref.  \cite{KKLT} and in
the model discussed above is that in  \cite{KKLT} the decay leads to
spontaneous decompactification of internal space in each bubble of the new
phase, whereas in the model proposed in this paper the tunneling to the space
with negative cosmological constant leads to the development of a cosmological
singularity inside each of the bubbles \cite{Coleman:1980aw}.

Note that dS space never decays completely. Just like in old inflation
\cite{Old} and in eternal inflation scenario \cite{Eternalnew,Eternalchaot},
the volume of its non-decayed parts will continue growing exponentially. In
some of its parts, the scalar field may jump upward to different minima of the
effective potential  \cite{HM},  supporting an eternal process of
self-reproduction of all possible vacuum states in stringy landscape.

In conclusion, in this paper we have found that the height of the barrier and the upper bound on the scale of inflation  in the simplest versions of the KKLT model are directly related to the present value of the gravitino mass. These observations could require development of particle phenomenology with large scale of SUSY breaking, or inflationary models with a very low scale of inflation, $H \lesssim m_{3/2}$. We suggested a modification of the original KKLT scenario where the volume stabilization does not require an uplifting of a deep AdS minimum, and where the large scale of inflation is compatible with the small gravitino mass.

It is a pleasure to thank C. Burgess, M. Dine, S. Kachru, P. Nilles, F. Quevedo, H. Stoica, S. Trivedi,  and  S.
Watson for useful discussions.  This work was supported in part by NSF grant
PHY-0244728.


\begin{thebibliography}{4}

\bibitem{Linde:1984je} A.D. Linde, ``The New Inflationary Universe Scenario,''
In: {\it  The Very Early Universe}, ed. G.W. Gibbons, S.W. Hawking and S.Siklos,  pp. 205-249 
(Cambridge University Press,1983);
A.~D.~Linde,
``Inflation Can Break Symmetry In SUSY,''
Phys.\ Lett.\ B {\bf 131}, 330 (1983);
A.~D.~Linde,
``The Inflationary Universe,''
Rept.\ Prog.\ Phys.\  {\bf 47}, 925 (1984).


\bibitem{Sakharov:1984ir}
A.~D.~Sakharov,
``Cosmological Transitions With A Change In Metric Signature,''
Sov.\ Phys.\ JETP {\bf 60}, 214 (1984)
[Zh.\ Eksp.\ Teor.\ Fiz.\  {\bf 87}, 375].


\bibitem{Lerche} W.~Lerche, D.~Lust and A.~N.~Schellekens,
 ``Chiral Four-Dimensional Heterotic Strings From Selfdual Lattices,''
Nucl. Phys. B {\bf 287}, 477 (1987).


\bibitem{Bousso} R.~Bousso and J.~Polchinski,
``Quantization of four-form fluxes and dynamical neutralization of the
cosmological constant,''
JHEP {\bf 0006}, 006 (2000)
[arXiv:hep-th/0004134].


\bibitem{KKLT}
S.~Kachru, R.~Kallosh, A.~Linde and S.~P.~Trivedi,
``De Sitter vacua in string theory,''
Phys.\ Rev.\ D {\bf 68}, 046005 (2003)
[arXiv:hep-th/0301240].

\bibitem{Silver}
E. Silverstein, ``(A)dS Backgrounds from Asymmetric Orientifolds,''
[arXiv:hep-th/0106209]; A. Maloney, E. Silverstein and A. Strominger,
``de Sitter Space in Noncritical String Theory,'' [arXiv:hep-th/0205316].





\bibitem{Susskind:2003kw}
L.~Susskind,
``The anthropic landscape of string theory,''
arXiv:hep-th/0302219;

\bibitem{Douglas:2003um}
M.~R.~Douglas,
``The statistics of string / M theory vacua,''
JHEP {\bf 0305}, 046 (2003)
[arXiv:hep-th/0303194]
M.~R.~Douglas, B.~Shiffman and S.~Zelditch,
``Critical points and supersymmetric vacua,''
arXiv:math.cv/0402326.
F.~Denef and M.~R.~Douglas,
``Distributions of flux vacua,''
JHEP {\bf 0405}, 072 (2004)
[arXiv:hep-th/0404116];
A.~Giryavets, S.~Kachru and P.~K.~Tripathy,
``On the taxonomy of flux vacua,''
JHEP {\bf 0408}, 002 (2004)
[arXiv:hep-th/0404243];
F.~Denef, M.~R.~Douglas and B.~Florea,
JHEP {\bf 0406}, 034 (2004)
[arXiv:hep-th/0404257].






\bibitem{Susskind:2004uv}
L.~Susskind,
``Supersymmetry breaking in the anthropic landscape,''
arXiv:hep-th/0405189.

\bibitem{Douglas:2004qg}
M.~R.~Douglas,
``Statistical analysis of the supersymmetry breaking scale,''
arXiv:hep-th/0405279;
M.~R.~Douglas, B.~Shiffman and S.~Zelditch,
``Critical points and supersymmetric vacua, II: Asymptotics and extremal
metrics,''
arXiv:math.cv/0406089;
M.~R.~Douglas,
``Basic results in vacuum statistics,''
Comptes Rendus Physique {\bf 5}, 965 (2004)
[arXiv:hep-th/0409207].


\bibitem{Dine:2004is}
M.~Dine, E.~Gorbatov and S.~Thomas,
``Low energy supersymmetry from the landscape,''
arXiv:hep-th/0407043.

\bibitem{Conlon:2004ds}
J.~P.~Conlon and F.~Quevedo,
``On the explicit construction and statistics of Calabi-Yau flux vacua,''
JHEP {\bf 0410}, 039 (2004)
[arXiv:hep-th/0409215].


\bibitem{Kachru}
O.~DeWolfe, A.~Giryavets, S. Kachru and W. Taylor, in preparation.



\bibitem{DeWolfe:2002nn}
O.~DeWolfe and S.~B.~Giddings,
``Scales and hierarchies in warped compactifications and brane worlds,''
Phys.\ Rev.\ D {\bf 67}, 066008 (2003)
[arXiv:hep-th/0208123].

\bibitem{Arkani-Hamed:2004fb}
N.~Arkani-Hamed and S.~Dimopoulos,
``Supersymmetric unification without low energy supersymmetry and signatures
for fine-tuning at the LHC,''
arXiv:hep-th/0405159; N.~Arkani-Hamed, S.~Dimopoulos, G.~F.~Giudice and A.~Romanino,
``Aspects of split supersymmetry,''
arXiv:hep-ph/0409232.

\bibitem{BKQ}
C.~P.~Burgess, R.~Kallosh and F.~Quevedo,
``de Sitter string vacua from supersymmetric D-terms,''
JHEP {\bf 0310}, 056 (2003)
[arXiv:hep-th/0309187].

\bibitem{Blanco-Pillado:2004ns}
J.J. Blanco-Pillado, C.P. Burgess, J.M. Cline, C. Escoda, M. Gomez-Reino, R. Kallosh, A. Linde , F. Quevedo,
``Racetrack inflation,''
arXiv:hep-th/0406230.




\bibitem{KKL} J.~P.~Hsu, R.~Kallosh and S.~Prokushkin,
``On brane inflation with volume stabilization,''
JCAP {\bf 0312}, 009 (2003)
[arXiv:hep-th/0311077];
J.~P.~Hsu and R.~Kallosh,
``Volume stabilization and the origin of the inflaton shift symmetry in string
theory,''
JHEP {\bf 0404}, 042 (2004)
[arXiv:hep-th/0402047];
K.~Dasgupta, J.~P.~Hsu, R.~Kallosh, A.~Linde and M.~Zagermann,
``D3/D7 brane inflation and semilocal strings,''
JHEP {\bf 0408}, 030 (2004)
[arXiv:hep-th/0405247];
S. Kachru, R. Kallosh and A. Linde, ``String Inflation Update,'' in preparation.

\bibitem{Linde:1988yp}
A.~D.~Linde and M.~I.~Zelnikov,
``Inflationary Universe With Fluctuating Dimension,''
Phys.\ Lett.\ B {\bf 215}, 59 (1988).

\bibitem{Buchmuller:2004xr}
W.~Buchmuller, K.~Hamaguchi, O.~Lebedev and M.~Ratz,
``Dilaton destabilization at high temperature,''
arXiv:hep-th/0404168.

\bibitem{Gen:2000nu}
U.~Gen and M.~Sasaki,
``Radion on the de Sitter brane,''
Prog.\ Theor.\ Phys.\  {\bf 105}, 591 (2001)
[arXiv:gr-qc/0011078].

\bibitem{Frolov:2003yi}
A.~V.~Frolov and L.~Kofman,
``Can inflating braneworlds be stabilized,''
Phys.\ Rev.\ D {\bf 69}, 044021 (2004)
[arXiv:hep-th/0309002];
C.~R.~Contaldi, L.~Kofman and M.~Peloso,
``Gravitational instability of de Sitter compactifications,''
JCAP {\bf 0408}, 007 (2004)
[arXiv:hep-th/0403270].


\bibitem{Navarro:2004mm}
I.~Navarro and J.~Santiago,
``Flux compactifications: Stability and implications for cosmology,''
JCAP {\bf 0409}, 005 (2004)
[arXiv:hep-th/0405173].

\bibitem{Giddings} S.~B.~Giddings,
``The fate of four dimensions,''
Phys.\ Rev.\ D {\bf 68}, 026006 (2003)
[arXiv:hep-th/0303031].

\bibitem{KKLMMT}
S.~Kachru, R.~Kallosh, A.~Linde, J.~Maldacena, L.~McAllister and
S.~P.~Trivedi, ``Towards inflation in string theory,'' JCAP {\bf
0310}, 013 (2003) [arXiv:hep-th/0308055].

\bibitem{GKP}
S.~B.~Giddings, S.~Kachru and J.~Polchinski, ``Hierarchies from
fluxes in string compactifications,'' Phys. Rev. {\bf D66}, 106006
(2002) [arXiv:hep-th/0105097].

\bibitem{Brustein:1992nk}
R.~Brustein and P.~J.~Steinhardt,
``Challenges for superstring cosmology,''
Phys.\ Lett.\ B {\bf 302}, 196 (1993) [arXiv:hep-th/9212049].

\bibitem{kaloper} N.~Kaloper and K.~A.~Olive,
``Dilatons in string cosmology,'' Astropart.\ Phys.\  {\bf 1}, 185 (1993);T.~Barreiro, B.~de Carlos and E.~J.~Copeland,
``Stabilizing the dilaton in superstring cosmology,'' Phys.\ Rev.\ D {\bf 58},
083513 (1998) [arXiv:hep-th/9805005]; A.~Linde,
``Creation of a compact topologically nontrivial inflationary universe,''
JCAP {\bf 0410}, 004 (2004)
[arXiv:hep-th/0408164]; R.~Brustein, S.~P.~de Alwis and P.~Martens,
Phys.\ Rev.\ D {\bf 70}, 126012 (2004)
[arXiv:hep-th/0408160];
 N.~Kaloper, J.~Rahmfeld and L.~Sorbo,
``Moduli entrapment with primordial black holes,''
arXiv:hep-th/0409226.


\bibitem{S}
R.~Kallosh and S.~Prokushkin,
``Supercosmology,''
arXiv:hep-th/0403060.

\bibitem{Kofman:2004yc}
L.~Kofman, A.~Linde, X.~Liu, A.~Maloney, L.~McAllister and E.~Silverstein,
``Beauty is attractive: Moduli trapping at enhanced symmetry points,'' JHEP
{\bf 0405}, 030 (2004) [arXiv:hep-th/0403001].

\bibitem{McAllister:2004gd}
L.~McAllister and I.~Mitra,
``Relativistic D-brane scattering is extremely inelastic,''
arXiv:hep-th/0408085.


\bibitem{W}
S.~Watson, ``Stabilizing moduli with string cosmology,'' arXiv:hep-th/0409281.

\bibitem{Weinberg:1982id}
S.~Weinberg,
``Does Gravitation Resolve The Ambiguity Among Supersymmetry Vacua?,''
Phys.\ Rev.\ Lett.\  {\bf 48} (1982) 1776.

\bibitem{Coleman:1980aw}
S.~R.~Coleman and F.~De Luccia,
``Gravitational Effects On And Of Vacuum Decay,''
Phys.\ Rev.\ D {\bf 21}, 3305 (1980);
T.~Banks,
``Heretics of the false vacuum: Gravitational effects on and of vacuum decay. II,''
arXiv:hep-th/0211160.

\bibitem{Old} A.~H.~Guth,
``The Inflationary Universe: A Possible Solution To The Horizon And Flatness
Problems,'' Phys.\ Rev.\ D {\bf 23}, 347 (1981); A.~H.~Guth and E.~J.~Weinberg,
``Could The Universe Have Recovered From A Slow First Order Phase Transition?,''
Nucl.\ Phys.\ B {\bf 212}, 321 (1983).

\bibitem{Eternalnew} A.~Vilenkin, ``The Birth Of Inflationary
Universes,'' Phys.\ Rev.\ D {\bf 27}, 2848 (1983).


\bibitem{Eternalchaot} A.~D.~Linde,
``Eternally Existing Self-reproducing Chaotic Inflationary Universe,''
Phys.\ Lett.\ B {\bf 175}, 395 (1986).

\bibitem{HM}
S.~W.~Hawking and I.~G.~Moss,
``Supercooled Phase Transitions In The Very Early Universe,''
Phys.\ Lett.\ B {\bf 110}, 35 (1982);
A.~A.~Starobinsky,
``Stochastic De Sitter (Inflationary) Stage In The Early Universe,''
in: {\sl Current Topics in Field Theory, Quantum Gravity
and Strings}, Lecture Notes in Physics, eds. H.J. de Vega and N. Sanchez
(Springer, Heidelberg 1986) {\bf 206}, p. 107;
A.~S.~Goncharov and A.~D.~Linde,
``Tunneling In Expanding Universe: Euclidean And Hamiltonian Approaches,''
Fiz.\ Elem.\ Chast.\ Atom.\ Yadra {\bf 17}, 837 (1986) (Sov. J. Part. Nucl.
{\bf 17}, 369 (1986)); K.~M.~Lee and E.~J.~Weinberg,
``Decay Of The True Vacuum In Curved Space-Time,''
Phys.\ Rev.\ D {\bf 36}, 1088 (1987); A.~D.~Linde,
``Hard art of the universe creation (stochastic approach to tunneling and
baby universe formation),''
Nucl.\ Phys.\ B {\bf 372}, 421 (1992)
[arXiv:hep-th/9110037];
A.~D.~Linde, D.~A.~Linde and A.~Mezhlumian,
``From the Big Bang theory to the theory of a stationary universe,''
Phys.\ Rev.\ D {\bf 49}, 1783 (1994)
[arXiv:gr-qc/9306035].
J.~Garriga and A.~Vilenkin,
``Recycling universe,''
Phys.\ Rev.\ D {\bf 57}, 2230 (1998)
[arXiv:astro-ph/9707292];
L.~Dyson, M.~Kleban and L.~Susskind,
``Disturbing implications of a cosmological constant,''
JHEP {\bf 0210}, 011 (2002)
[arXiv:hep-th/0208013];
B.~Freivogel and L.~Susskind,
``A framework for the landscape,''
Phys.\ Rev.\ D {\bf 70}, 126007 (2004)
[arXiv:hep-th/0408133];
S.~M.~Carroll and J.~Chen,
``Spontaneous Inflation and the Origin of the Arrow of Time,''
arXiv:hep-th/0410270.


\end{thebibliography}
\end{document}